\documentclass[10pt,letterpaper]{article}
\usepackage[top=0.85in,left=2.75in,footskip=0.75in]{geometry}

\usepackage{amsmath,amssymb}

\usepackage{changepage}

\usepackage[utf8x]{inputenc}

\usepackage{textcomp,marvosym}

\usepackage{cite}

\usepackage{nameref,hyperref}

\usepackage[right]{lineno}

\usepackage{microtype}
\DisableLigatures[f]{encoding = *, family = * }

\usepackage[table]{xcolor}

\usepackage{array}

\newcolumntype{+}{!{\vrule width 2pt}}

\newlength\savedwidth

\raggedright
\usepackage[aboveskip=1pt,labelfont=bf,labelsep=period,justification=raggedright,singlelinecheck=off]{caption}

\makeatletter
\renewcommand{\@biblabel}[1]{\quad#1.}
\makeatother

\usepackage{lastpage,fancyhdr,graphicx}
\usepackage{epstopdf}
\fancyheadoffset[L]{2.25in}
\fancyfootoffset[L]{2.25in}
\begin{document}
\vspace*{0.2in}

\begin{flushleft}
{\Large
\textbf\newline{Optimal periodic closure for minimizing risk in emerging disease outbreaks} % Please use "sentence case" for title and headings (capitalize only the first word in a title (or heading), the first word in a subtitle (or subheading), and any proper nouns).
}
\newline
\\
Jason Hindes\textsuperscript{1},
Simone Bianco\textsuperscript{2}, 
and Ira B. Schwartz\textsuperscript{1}
\\
\bigskip
\textbf{1} U.S. Naval Research Laboratory, Washington, DC 20375, USA
\\
\textbf{2} IBM Almaden Research Center, San Jose, CA 95120-6099, USA
\\
\bigskip

* jason.hindes@nrl.navy.mil

\end{flushleft}
\section*{Abstract}
Without vaccines and treatments, societies must rely on non-pharmaceutical
intervention strategies to control the spread of emerging diseases such as
COVID-19. Though complete lockdown is epidemiologically effective, because
it eliminates infectious contacts, it comes with significant costs. Several recent studies have suggested that a plausible compromise strategy
for minimizing epidemic risk is periodic closure, in which populations oscillate between wide-spread social restrictions and relaxation. 
However, no underlying theory has been proposed to predict and explain optimal closure periods as a function
of epidemiological and social parameters. In this work we develop such an analytical theory for SEIR-like model diseases, 
showing how characteristic closure periods emerge that minimize the total outbreak, and increase
predictably with the reproductive number and incubation periods of a disease-- as long as both 
are within predictable limits. Using our approach we demonstrate a sweet-spot effect in which optimal periodic closure
is maximally effective for diseases with similar incubation and recovery
periods. Our results compare well to numerical simulations, including in COVID-19 models
where infectivity and recovery show significant variation.

% Please keep the Author Summary between 150 and 200 words
% Use first person. PLOS ONE authors please skip this step. 
% Author Summary not valid for PLOS ONE submissions.   
\section*{Author summary}
The COVID-19 pandemic has revealed how emergent infectious diseases can be extremely disruptive 
to a society, both in terms of socio-economic and health outcomes. In the early days of an emergent disease pandemic, the primary method for societal-level control 
revolves around reducing contacts between people, which in the most extreme cases takes the form 
of complete lockdown and stay-at-home orders. However, for obvious economic and social reasons, indefinitely long 
lock-down has proven hard to maintain. Consequently, many recent modeling and simulation efforts have focused on 
determining optimal cycles of closure and relaxation. In this work, we develop an analytical approach for 
predicting periodic-closure cycles that minimize the size of disease outbreaks. We show how a characteristic optimal period arises, 
for diseases with tunable periodic contact rates, as a function of the reproductive number and incubation periods of a disease.

\linenumbers
\begin{nolinenumbers}
% Use "Eq" instead of "Equation" for equation citations.
\section{Introduction}
The COVID19 pandemic, caused by the novel RNA
virus SARS-CoV-2~\cite{Wang2020a}, has resulted in  
devastating health, economic, and social consequences.
In the absence of vaccines and treatments, non-pharmaceutical intervention (NPI) strategies
have been adopted to varying degrees around the world. 
Given the nature of the virus transmission, NPI measures have 
effectively reduced human contacts-- both slowing the pandemic, and minimizing the risk of local outbreaks~\cite{Ferguson2020,Wang2020}.  
The use of drastic NPI strategies in China reportedly reduced
the basic reproductive number, $R_0$, to a value smaller than $1$, strongly
curbing the epidemic within a short period of time ~\cite{Wang2020,Pan2020}.  
On the other hand widespread testing protocols and contact tracing, in e.g., South Korea, 
significantly controlled spread during the initial phase of the pandemic~\cite{Cohen2020}. 
%In addition, where rigorous and widespread testing protocols have been adopted,
%such as South Korea, the spread has been controlled~\cite{Cohen2020}
In other countries, the implementation of NPI policies has not been as
strict~\cite{Ferguson2020}, with an optimistic reduction in transmission of 
roughly a half. \textcolor{black}{To complicate the containment of the disease, early reports indicated 
significant amounts of pre-symptomatic and asymptomatic transmission~\cite{Li2020,LiGuan2020}. For instance, recent estimates point to 
asymptomatic infection accounting for around $20\!-\!30\%$ of the total, with a similar percentage for 
pre-symptomatic infections\cite{Moghadas17513}-- together producing a majority. These findings have been supported by other experimental studies \cite{Sutton2020_UniversalScreening} and analysis of the existing data~\cite{Levin2020,Schwartz2020}.} 

%\ira{The potential of COVID19 as well as other novel viruses to spread unchecked
%puts a strong pressure on the rapid design and deployment of controls such
%as social distancing and testing. Where rigorous and widespread testing protocols have been adopted,
%such as South Korea, the spread has been controlled~\cite{Cohen2020}. An
%optimal control release strategy may include testing individuals before social
%distancing release~\cite{Schwartz2020}.}

As NPI controls such as quarantine, social distancing and testing are enforced, it is important to understand the impact
of early release and relaxation of controls on the affected populations\cite{ZHAO2020108405,Serra2020}.   
Recent studies have attempted to address how societies can vary
social contacts optimally in time in order to maintain economic activity while controlling
epidemics~\cite{Karin2020cyclic}. For instance, preliminary numerical studies
suggest that periodic closure to control outbreak risk, where a population
oscillates between 30-50 days of strict lockdown followed by 30-50 days of
relaxed social restrictions, may efficiently contain the spread of COVID-19
and minimize economic damage~\cite{Chowdhury2020dynamic}. These studies test interesting hypotheses, but cannot be immediately generalized to new emerging diseases. A basic
understanding of why and when such risk minimizing strategies are effective remains unclear, and may benefit from a general analytical approach.

As a first step in this direction we analyze SEIR-like models with tunable
periodic contact rates. Our methods reveal the
existence of a characteristic optimal period of contact-breaking between
individuals that minimizes the risk of observing
a large outbreak, and predicts exactly 
how such an optimal period depends on 
epidemic and social parameters. In particular, we show that the optimal period for closure increases (or decreases) predictably
with $R_0$ and the incubation period of a disease, and exists as long as $R_0$ is below 
a predictable threshold, and when there is not a time-scale separation between 
incubation and recovery. We demonstrate analytically that periodic closure is maximally effective for containing 
disease outbreaks when the typical incubation and recovery periods for a disease are similar -- in such cases suppressing large outbreaks with $R_0$'s as large as $4$.
Our results compare well to numerical simulations and are robust to the inclusion of heterogeneous infection and recovery rates, 
which are known to be important for modeling COVID-19 dynamics. 

To begin, we first consider the canonical SEIR model with a time-dependent infectious contact rate parameter,
$\beta(t)$. Individuals in this model are in one of four possible states:
susceptible, exposed, infectious, and recovered. Following the simplest
mass-action formulation of the disease dynamics, and assuming negligible background births
and deaths, the fraction of susceptible $(s)$, exposed $(e)$, infectious $(i)$, and recovered $(r)$ individuals in a population satisfy the following differential equations in time (t), where dots denote time derivatives:
\begin{align}
\label{eq:SEIR}
\dot{s}=\;&-\beta(t)si, \\ 
\dot{e}=\;&\beta(t)si -\alpha e,\\
\dot{i}=\;&\alpha e -\gamma i,\\
\label{eq:SEIRf}
\dot{r}=\;&\gamma i.
\end{align}

Such equations are valid in in the limit of large, well-mixed populations and
constitute a baseline description for the spreading of many
diseases\cite{aron1984seasonality,Keeling:Book}. Note that $\alpha$ is the
rate at which exposed individuals become infectious,
while $\gamma$ is the rate at which infected individuals recover. If
$\beta(t)\!=\!\beta_{0}=~$constant, it is straightforward to show that the
basic reproductive number for the SEIR model, $R_{0}$, which measures the
average number of new infections generated by a single infectious individual in a fully susceptible population, is $R_{0}\!=\!\beta_{0}/\gamma$ \cite{Hefferman2005,Keeling:Book,AndersonAndMay}. Note in this work when $R_{0}$ is written as a constant (no time dependence) it should be taken to mean this value. Typical values for the $R_{0}$ of COVID-19 range from $1\!-\!4$, depending on local population contact rates\cite{Covasim,Pan2020}.

\section{\label{sec:methods} Methods}
As a simple model for periodic closure we assume a step function for $\beta(t)$ with infectious contacts occurring for a period of $T$ days with rate $\beta_{0}$, followed by no contacts for the same period, $\beta(t)=\beta_{0}\cdot\text{mod}({\text{floor}}\{[t+T]/T\},2)$~\cite{Floornote}. A schematic of $\beta(t)$ is plotted in the inlet panel of Fig.~\ref{fig:TimeSeries}(a). \textcolor{black}{In \nameref{S1_Appendix} we show results for smoothly varying $\beta(t)$ and asymmetric closure, where lockdown and open contacts occur for different amounts of time. It is demonstrated that the results presented in the main text do not qualitatively change under these generalizations.} Also in Fig.~\ref{fig:TimeSeries}(a), we plot an example time-series of the infectious fraction, normalized by the initial fraction of non-susceptibles, for three different closure periods: green (short), blue (intermediate), and red (long). For periods that are not too long or short, the disease remains in a linear spreading regime (as we will show below), and therefore normalizing by the initial conditions gives time series that are initial-condition independent.    
\begin{figure}[h]
\center{\includegraphics[scale=0.26]{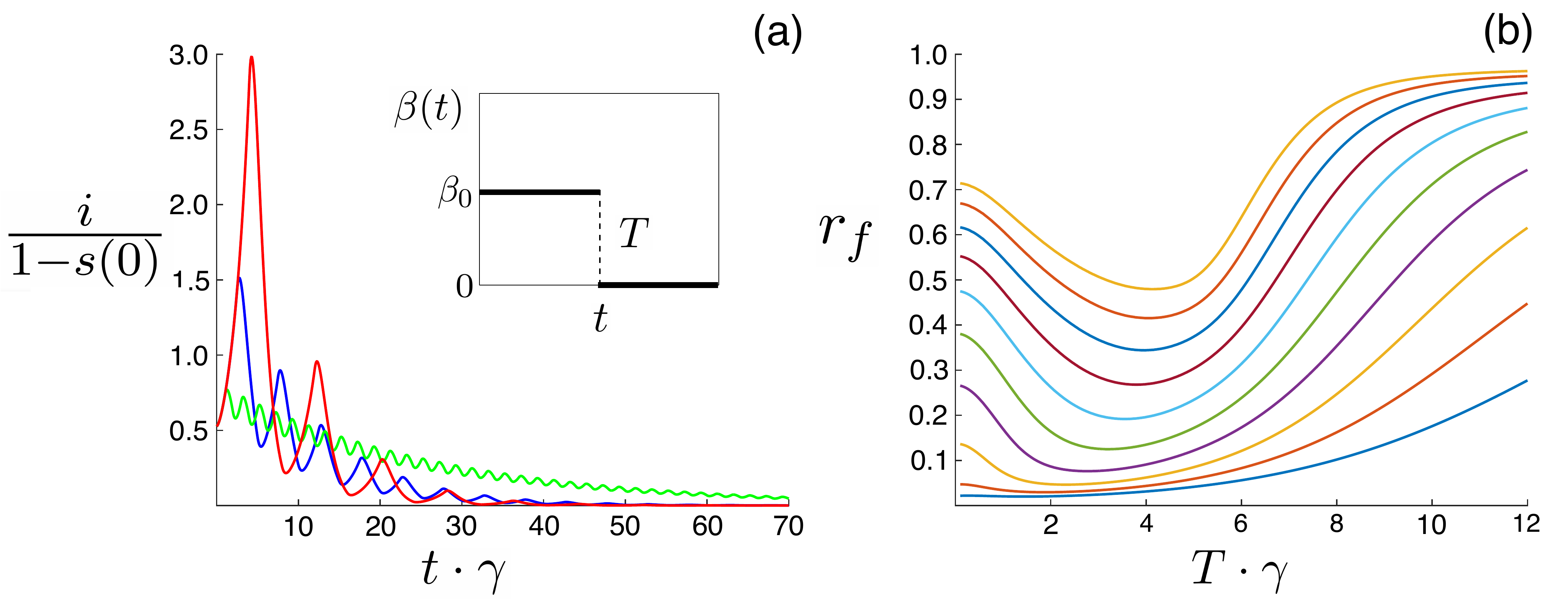}}
\caption{Periodic closure examples. (a) fraction infectious, normalized by initial conditions, versus time for $T\!=\!10\!\cdot\!\text{days}$ (green), $T\!=\!25\!\cdot\!\text{days}$ (blue), $T\!=\!40\!\cdot\!\text{days}$ (red) closure periods. The inlet panel shows a schematic of $\beta(t)$. Other model parameters are: $\gamma^{-1}\!=\!10\cdot\!\text{days}$, $\alpha^{-1}\!=\!8.33\cdot\!\text{days}$, and $\beta_{0}^{-1}\!=\!5\cdot\!\text{days}$. (b) Outbreak size versus the closure period. Curves correspond to different $R_{0}\!=\!\beta_{0}/\gamma$, starting from the bottom: first $(R_{0}\!=\!1.5)$, second $(R_{0}\!=\!1.7)$, ..., top $(R_{0}\!=\!3.3)$. Other model parameters are identical to (a).}
\label{fig:TimeSeries}
\end{figure} 

Intuitively, since the incubation period, $\alpha^{-1}$, is finite, it takes
time to build-up infection from small initial values. As a consequence, we
expect that it may be possible to allow some disease exposure, before cutting
contacts, and the result may be a net reduction in infection at the end of a
closure period. For instance, notice that all $i(t)$ decrease over a full
closure cycle, $2T$, in Fig.~\ref{fig:TimeSeries}(a). If the closure period is
too small, infection can still grow (e.g., as $T\!\rightarrow\!0$,
$R_{0}(t)\!\sim\!\left<R_{0}(t)\right>_{t}\!=
\!R_{0}/2$ which could be above the epidemic
threshold), while if the period is too long, a large outbreak
will occur before the control is applied. Between these two limits, there should be
an optimal $T$ $(T_{\text{min}})$, that results in a minimum outbreak. To
illustrate, in Fig.~\ref{fig:TimeSeries}(b) we show an example of the final
outbreak-size, $r(t\!\rightarrow\!\infty)\equiv\!r_{f}$ starting from $i(t=0)\!=\!10^{-3}$, as a function of the
closure period for different, equally spaced values of $R_{0}$: the bottom curves
correspond to smaller values of $R_{0}$, while the top curves
correspond to larger values. 

As expected from the above intuitive argument, simulations show an optimal period that minimizes $r_{f}$. A natural question is, how does $T_{min}$ depend on model parameters? Our approach in the following is to develop theory for $T_{\text{min}}$ in the SEIR-model, and then show how such a theory can be easily adapted to predict $T_{\text{min}}$ in more complete models, e.g., in COVID-19 models that include heterogeneous infectivity and asymptomatic spread\cite{Covasim,Schwartz2020}. 

\section{Results}
\subsection{\label{sec:Anal}Optimal control} 
It is possible to estimate $T_{\text{min}}$ by calculating its value in the
linearized SEIR model, applicable when the fraction of non-susceptibles is
relatively small. When $e(t), i(t), r(t),1\!-\!s(t)\!\ll\!1$, the dynamics of Eqs.(\ref{eq:SEIR}-\ref{eq:SEIRf}) are effectively driven by a 2-dimensional system:
\begin{align}
\label{eq:LinearSystem1}
\frac{d\boldsymbol{\Psi}}{dt}&=\gamma\bold{M}(t)\!\cdot\!\boldsymbol{\Psi},\\
\label{eq:LinearSystem2}
\bold{M}(t)&= \begin{bmatrix}
-a & R_{0}(t) \\
\;\;\;a & -1
\end{bmatrix}, 
\end{align}
where $a\!\equiv\!\alpha/\gamma$, $R_{0}(t)\!\equiv\!\beta(t)/\gamma$, and
$\boldsymbol{\Psi}(t)^\top\!=\![e(t),i(t)]$. 

The first step in calculating $T_{\text{min}}$ is to construct eigen-solutions of Eqs.(\ref{eq:LinearSystem1}-\ref{eq:LinearSystem2}) in the form 
\begin{align}
\label{eq:Floquet}
\boldsymbol{\Psi}^{p}(2T)=\nu(T)\cdot\boldsymbol{\Psi}^{p}(0),
\end{align}
where $\nu(T)$ is the largest such eigenvalue; the superscript $p$ denotes the corresponding principal eigenvector. Ignoring the subdominant
eigenvalues assumes that after a sufficiently large number
of iterations of periodic closure, the dynamics is well aligned with the principle solution no matter what the initial conditions. Unless stated otherwise, simulations are started in this state so that initial-condition effects are minimized. The second step is to calculate the integrated incidence, $r(2T)$ from the solution of Eq.~\eqref{eq:Floquet}, by integrating $i(t)$ over a full cycle
\begin{align}
\label{eq:r2T}
r(2T)=\int_{0}^{2T}\![\bold{\Psi}^{p}(t)]_{2}\cdot \gamma dt, 
\end{align}
where $[\bold{\Psi}^{p}(t)]_{2}$ denotes the infectious-component of $\boldsymbol{\Psi}^{p}(t)$. The third step is to calculate the final outbreak size from $r(2T)$. To this end, it is important to realize that as long as $\nu(T)\!<\!1$, the outbreak will decrease {\it geometrically} after successive closure cycles, and therefore $r_{f}(T)\!=\!r(2T)\!+\!\nu(T)r(2T)\!+\!\nu(T)^{2}r(2T)+...$, or
\begin{align}
\label{eq:Final}
r_{f}(T) = r(2T)/[1-\nu(T)].  
\end{align} 
Finally, we can find the local minimum of $r_{f}(T)$ when $\nu(T)\!<\!1$ by solving 
\begin{align}
\label{eq:MinCondition}
\frac{dr_{\!f}}{\!dT}\Big\vert_{T_{\text{min}}}=0. 
\end{align}
This algorithm gives a single fixed-point equation that determines $T_{\text{min}}$.         

Since our analysis is based on a piecewise 2-dimensional linear system,
it is possible to give every quantity in the previous paragraph an exact
expression\cite{StrogatzBook} in terms of epidemiological and social parameters. See \nameref{S1_Appendix} for full derivation and exact expressions for Eqs.(\ref{eq:Floquet}-\ref{eq:MinCondition}). Following our procedure gives the prediction curves shown in Fig.~\ref{fig:OptimalClosure}(a). The solid red line indicates the solution to Eq.~\eqref{eq:MinCondition}, and agrees well with simulation-determined minima of $r_{f}(T)$ over a range of $R_{0}$ given initial fractions of infectious $10^{-6}$ (circles), $10^{-4}$ (squares), and $10^{-2}$ (diamonds). The simulation-determined minima are computed from $r_{f}(T)$ curves like Fig.\ref{fig:TimeSeries}(b). \textcolor{black}{It is important to note that our optimal-control theory assumes the validity of the linearized SEIR model, applicable when the total outbreak size, $r_{f}\!\ll\!1$. In general, the total outbreak size will increase with the initial fraction of infectious and $R_{0}$, and hence, the larger both are, the more simulations will disagree with theory. For example, this explains the better agreement for initial fractions of infectious $10^{-6}$, as compared to $10^{-2}$ in Fig.~\ref{fig:OptimalClosure}(b).} 

On the other hand, the solid blue line in Fig.\ref{fig:OptimalClosure}(a) indicates the threshold closure period, satisfying
\begin{align}
\label{eq:Thresh}
\nu(T_{\text{thresh}})=1. 
\end{align}
\textcolor{black}{The closure period $T_{\text{thresh}}$ results in the largest eigenvalue of Eqs.(\ref{eq:LinearSystem1}-\ref{eq:LinearSystem2}) equalling unity such that the principal component of exposed and infectious fractions is unchanged after a full closure cycle.
If $T\!<\!T_{\text{thresh}}$, $\nu(T)\!>\!1$ and a large outbreak occurs, even with closure, as infection grows over a full cycle for any small non-zero $\boldsymbol{\Psi}(0)$.} Given this property, $T_{\text{thresh}}$ gives a lower bound for the optimal period, $T_{\text{min}}\!>\!T_{\text{thresh}}$. Note: the red curve is always above the blue curve in Fig.\ref{fig:OptimalClosure}(a).   
\begin{figure}[h]
\center{\includegraphics[scale=0.205]{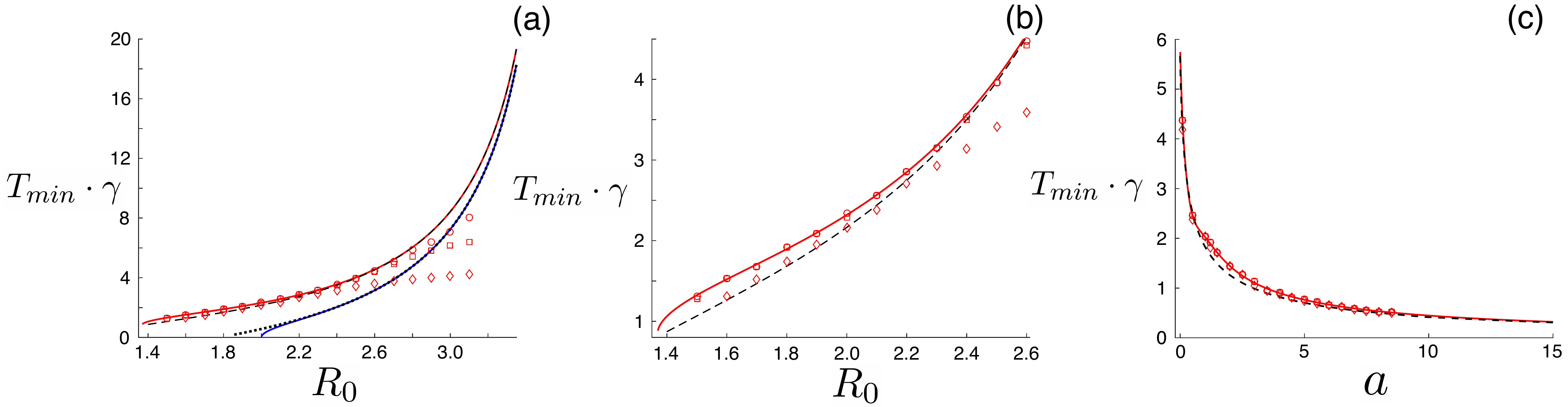}}
%\center{\includegraphics[scale=0.245]{Fig_2_PeriodicClosure.pdf}}
\caption{Optimal periodic closure. (a) Period versus $R_{0}\!=\!\beta_{0}/\gamma$. The solid-red  and dashed lines are theoretical predictions (exact and approximate, respectively), and the points are simulation-determined minima for initial fractions infectious: $10^{-6}$ (circles), $10^{-4}$ (squares), and $10^{-2}$ (diamonds). The blue and dotted curves are predictions for the threshold closure period (exact and approximate, respectively). Other model parameters are: $\gamma^{-1}\!=\!10\!\cdot\!\text{days}$ and $\alpha^{-1}\!=\!8.33\!\cdot\!\text{days}$. \textcolor{black}{(b) A refocused version of (a) for smaller values of $R_{0}$.} (c) Period versus $a\!=\!\alpha/\gamma$. The color scheme and parameters are identical to (a), except $\beta^{-1}\!=\!5.55\!\cdot\!\text{days}$.}
\label{fig:OptimalClosure}
\end{figure}

Before analyzing Eqs.(\ref{eq:LinearSystem1}-\ref{eq:MinCondition}) further, we point out two basic dependencies in the (normalized) optimal period
$T_{min}\!\cdot\gamma$. The first is intuitive: as the reproductive number 
$R_{0}$ increases, so does $T_{min}\cdot\gamma$. Hence, the faster a disease
spreads the longer a population's closure-cycle must be in order to contain
it. The second is more interesting. Notice in Fig.~\ref{fig:OptimalClosure}(c)
that $T_{min}\cdot\gamma\rightarrow\!\infty$ as $a\!\rightarrow\!0$, and
$T_{min}\cdot\gamma\!\rightarrow\!0$ as $a\!\rightarrow\!\infty$. Therefore, recalling $a\!=\!\alpha/\gamma$,
if a disease has a long incubation period, then the optimal closure cycle is
similarly long. On the other hand, if a disease has a short incubation period,
then the optimal closure cycle is short. In order for periodic
closure to be a practical strategy, with {\it a finite} $T_{min}$, our results indicate that $a\!\sim\!\mathcal{O}(1)$, roughly speaking, or that the recovery
and incubation periods should be on the same time scale-- a condition that generally applies to acute infections\cite{AndersonAndMay}. 

Another observation from our approach that we can make is that periodic closure is not an
effective strategy for arbitrarily large $R_{0}$, as one might expect. One
way to see this from the analysis is to notice that the {\it optimal period
diverges for the linear system} at some $R_{0}^{\text{max}}$, as
$T_{\text{thresh}}\!\rightarrow\!T_{\text{min}}\!\rightarrow\!\infty$ (at
fixed $a$). This transition can be seen in
Fig.\ref{fig:OptimalClosure}(a), as the blue and red curves collide. Above the
transition $R_{0}\!>\!R_{0}^{\text{max}}$, no periodic closure can keep a
disease from growing over a cycle. In this sense
$R_{0}^{\text{max}}(a)$ gives an upper bound on contact rates between individuals that can be suppressed by periodic-closure as a control strategy. We note that an optimal $T_{\text{min}}$ still exists even when our linear approximation no longer applies, e.g., $R_{0}\!>\!R_{0}^{\text{max}}$ (in the sense that $r(t\rightarrow\infty)$ is minimized by some $T_{\text{min}}$), but the benefit of control becomes smaller and smaller as $R_{0}$ is increased, and the optimal period becomes increasingly dependent on initial conditions. In such cases, one must resort to numerical simulations of the full non-linear system, Eqs.(\ref{eq:SEIR}-\ref{eq:SEIRf}). 

A sharper analytical understanding can be found by making the additional
approximation that $\boldsymbol{\Psi}(t)\!\sim\!\exp[\lambda_{11}\gamma t]\bold{v}_{11}$,
for $t\!<\!T$ and $\beta(t)\!=\!\beta_{0}$, where
\begin{align}
\label{eq:L11}
\lambda_{11}=\frac{-a-1+\sqrt{(a+1)^{2}+4a(R_{0}-1)}}{2}. 
\end{align}
Equation (\ref{eq:L11}) is the largest eigenvalue of ${\bf M}(t\!<\!T)$ with eigenvector $\bold{v}_{11}$.
Hence, we ignore the time-decaying part, $\boldsymbol{\Psi}(t)_{\text{dec}}\!\sim\!\exp[-(a+1+\lambda_{11})\gamma t\;]\bold{v}_{12}$, of a
general solution where $\bold{v}_{12}$ is the other eigenvector of ${\bf M}(t\!<\!T)$. Our assumption becomes increasingly accurate with increasing
$T$, and Eqs.(\ref{eq:Floquet}-\ref{eq:Thresh}) simplify significantly:
\begin{align}
\label{eq:NuA}
&\nu(T)\approx e^{T\gamma\lambda_{11}}\Big[\;fe^{-T\gamma} +(1-f)e^{-Ta\gamma}\;\Big],
\end {align}
\begin{align}
\label{eq:R2A}
&\frac{r(2T)}{\bar{r}}\approx\frac{e^{T\gamma\lambda_{11}}\!-\!1}{\lambda_{11}} \;\;\;+ \nonumber \\
&\frac{e^{T\gamma\lambda_{11}}}{1-a}\!\Bigg(\!\!\frac{(\lambda_{11}\!+\!1)(1\!-\!e^{-T\gamma a})}{a} \!-(a\!+\!\lambda_{11})(1\!-\!e^{-\gamma T})\!\!\Bigg),
\end{align}
where
\begin{align}
\label{eq:p}
f= \frac{(\lambda_{11}+a)^{2}}{(a-1)(2\lambda_{11}+a+1)},  
\end{align}
and $\bar{r}$ is a constant that depends on $\beta_{0}, \;\alpha, \;\gamma$ and initial conditions, but is independent of $T$. Substituting Eqs.(\ref{eq:NuA}--\ref{eq:p}) into Eqs.(\ref{eq:MinCondition}--\ref{eq:Thresh}) gives a single fixed-point equation for the approximate $T_\text{min}$ and $T_\text{thresh}$ each, which can be easily solved. See \nameref{S1_Appendix} for further details. Examples of the approximate solutions are plotted with dotted and dashed lines in Fig.~\ref{fig:OptimalClosure}, and are almost indistinguishable from the complete linear-theory predictions shown with solid lines.

Using the simplified expressions, we can now show several interesting features of periodic closure. First, since Eqs.(\ref{eq:NuA}-\ref{eq:R2A}) are exact for large $T$, we can determine $R_{0}^{\text{max}}$ as a function of $a$. As $T\!\rightarrow\!\infty$, Eq.(\ref{eq:NuA}) has two scaling limits depending on whether $a\!\ge\!1$ or $a\!<\!1$. In the former, the second term on the RHS of Eq.(\ref{eq:NuA}) becomes negligible. As $T\!\rightarrow\!\infty$ the solution of $\nu\!=\!1$  is   
$\lambda_{11}\!\rightarrow\!1$. Solving for $R_{0}$ in $\lambda_{11}\!=\!1$ gives $R_{0}^{\text{max}}$. Similarly when $a\!<\!1$, as $T\!\rightarrow\!\infty$ the solution of $\nu\!=\!1$ is $\lambda_{11}\!\rightarrow\!a$. Putting the two cases together, gives $R_{0}^{\text{max}}(a)$, and the phase-diagram for optimal-periodic closure:
\begin{equation}
\label{eq:betamax}
R_{0}^{\text{max}} = \begin{cases}1+ (a+2)/a &\text{if }a \ge 1, \\ 
 2\big(a+1\big) &\text{if }a< 1. \end{cases}
\end{equation}
Equation (\ref{eq:betamax}) is plotted in Fig.\ref{fig:BetaMax}. In region I, the optimal period is predicted to be finite, in which case small outbreaks can be contained by optimal closure. In region II, such outbreaks can not be contained.
\textcolor{black}{The blue squares plot numerically-determined thresholds for the piecewise linear system Eqs.(\ref{eq:LinearSystem1}-\ref{eq:Floquet}) in the long closure-time limit. We compute each point by: picking a fixed value of $a$ (starting with $R_{0}\!=\!2$), solving for $T_{\text{thresh}}$ according to Eqs.(\ref{eq:LinearSystem1}-\ref{eq:Floquet}) and Eq.(\ref{eq:Thresh}), and then repeatedly incrementing $R_{0}$ in small steps of $0.001$ and solving for $T_{\text{thresh}}(R_0,a)$ until it is a large number, i.e., $T_{\text{thresh}}(R_{0}, a)\!\cdot\!\gamma\!=\!500$. Note that as long as the system Eqs.(\ref{eq:SEIR}-\ref{eq:SEIRf}) is below threshold, we can always start with initial fractions of infectious and exposed that are small enough for the linear system to apply.}     
\begin{figure}[h]
%\center{\includegraphics[scale=0.285]{Fig_3_PeriodicClosure.pdf}}
\center{\includegraphics[scale=0.285]{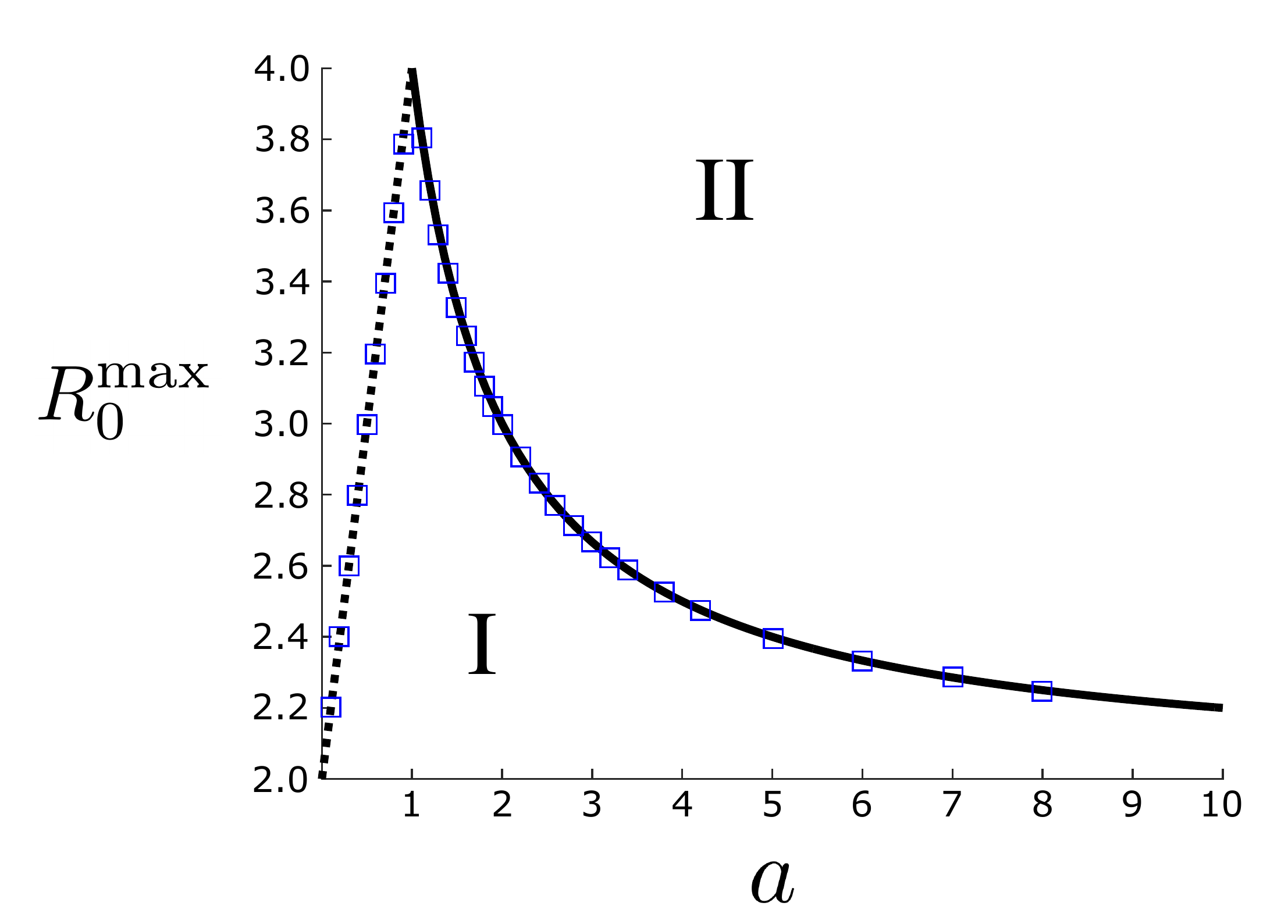}}
\caption{The largest reproductive number $R_{0}$ for which periodic closure can keep an SEIR-model disease under threshold. The two regimes are $a=\alpha/\gamma\!\ge\!1$ (solid line) and $a\!<\!1$ (dashed line). \textcolor{black}{Blue squares represent the numerically-determined threshold for the piecewise linear system Eqs.(\ref{eq:LinearSystem1}-\ref{eq:Floquet}) in the long closure-time limit.} In region I, outbreaks are contained by optimal closure. In region II, they are not.}
\label{fig:BetaMax}
\end{figure}

There are several important cases to notice in Fig.\ref{fig:BetaMax}. The first is that $R_{0}^{\text{max}}$ has a peak when $a\!=\!1$ ($\alpha\!=\!\gamma$). The implication is that periodic closure has the largest range of effectiveness, as measured by the ability to keep infection from growing over any closure-cycle, for diseases with {\it equal} exposure and recovery times. In this symmetric case, periodic closure can prevent large outbreaks as long as $R_{0}\!<\!4$ (compare this to the usual epidemic threshold without closure, $R_{0}\!=\!1$). On the other hand, when there is a time-scale separation between incubation and recovery, $a\!\rightarrow\!\infty$ or  $a\!\rightarrow\!0$, the phase-diagram nicely reproduces the
intuitive, time-averaged effective epidemic threshold $\left<R_{0}(t)\right>_{t}=\!1$, or $R_{0}^{\text{max}}=2$. 

\subsection{\label{sec:Hetero} COVID-19 model} 
Now we turn our attention to more complete models that derive from the basic SEIR-model assumptions, but have more disease classes and free parameters which are necessary for accurate predictions. In particular, epidemiological predictions for COVID-19 seem to require an asymptomatic disease state, i.e., a group of people capable of spreading the disease without documented symptoms. Such asymptomatic transmission is thought to be a significant driver for the worldwide distribution of the disease\cite{Chinazzi2020,Lavezzo2020}, since symptomatic individuals can be easily identified for quarantining while asymptomatics cannot (without widespread testing). Many models have been proposed to incorporate the broad spectrum of COVID-19 symptoms, as well as control strategies such as testing-plus-quarantining\cite{Covasim,Schwartz2020}. A common feature of such models is the assumption that exposed individuals enter into one of several possible infectious states according to a prescribed probability distribution (e.g., asymptomatic, mild, severe, tested-and-infectious, etc.) with their own characteristic infection rates and recovery times. Following this general prescription, we define $M$ infectious classes, $i_{m}$, where $m\!\in\!\{1,2,...M\}$, each with its own infectious contact rate $\beta_{m}(t)$ and recovery $\gamma_{m}$ rate, and which appear  
from the exposed state with probabilities $p_{m}$. The relevant {\it heterogeneous} SEIR-model equations become 
\begin{align}
\label{eq:SEIRnew1} 
\frac{de}{dt}=\;&\sum_{m}\beta_{m}(t)i_{m}s -\alpha e,\\
\label{eq:SEIRnew2}
\frac{di_{m}}{dt}=\;&\alpha p_{m} e -\gamma_{m} i_{m}.
\end{align}

Taking a common closure cycle for all individuals in the population, $\beta_{m}(t)=\beta_{0,m}\cdot\text{mod}({\text{floor}}\{[t+T]/T\},2)$~\cite{Floornote}, we would like to test our method for predicting $T_{\text{min}}$ in the more general model Eqs.(\ref{eq:SEIRnew1}-\ref{eq:SEIRnew2}), and demonstrate robustness to heterogeneity. In terms of an algorithm, we could simply repeat our approach for the effective $1+M$ dimensional linear system; though, we loose  
analytical tractability. On the other hand, because $T_{\text{min}}$ is well captured by a linear theory, which depends only on $R_{0}$, $a$, and $\gamma$, we might guess that quantitative accuracy can be maintained for higher dimensional models such as Eqs.(\ref{eq:SEIRnew1}-\ref{eq:SEIRnew2}) by swapping in suitable values for these parameters in our SEIR-model formulas above. This is analogous to the epidemic-threshold condition ($R_{0}\!=\!1$) being maintained in such models, as long as the correct value of $R_{0}$ is assumed. 

The $R_{0}$ for Eqs.(\ref{eq:SEIRnew1}-\ref{eq:SEIRnew2}) is easy to derive using standard methods\cite{Hefferman2005,Keeling:Book}, 
\begin{align}
\label{eq:R0new1} 
R_{0}=\sum_{m} p_{m}\beta_{0,m}/\gamma_{m}.  
\end{align}
Note: the updated $R_{0}$ is simply an average over the reproductive numbers for each infectious class. Using this averaging pattern as a starting point, our approach is to substitute the average values of $\alpha/\gamma_{m}$ and $\gamma_{m}$,  
\begin{align}
\label{eq:Anew1} 
a=\sum_{m} p_{m}\alpha/\gamma_{m} 
\end{align} 
\begin{align}
\label{eq:Gnew1} 
\gamma=\sum_{m} p_{m}\gamma_{m},   
\end{align}
into Eqs.(\ref{eq:Floquet}-\ref{eq:MinCondition}), or Eqs.(\ref{eq:NuA}-\ref{eq:p}) for approximate solutions. Namely, for the SEIR model we have an equation $0\!=\!F(R_{0},a,\gamma,T_{min})$, where $F$ is a function that is determined from Eq(\ref{eq:MinCondition}). Our averaging approximation entails solving the same Eq.(\ref{eq:MinCondition}) for $T_{\text{min}}$, but with parameters given by Eqs.(\ref{eq:R0new1}-\ref{eq:Gnew1}).  

We point out that this approximation is not arbitrary since in the limit of heterogeneous infectivity only, $\gamma_{m}\!=\!\gamma\;\forall m$, one solution of Eqs.(\ref{eq:SEIRnew1}-\ref{eq:SEIRnew2}) is $i_{m}(t)\!=\!p_{m}i(t)$, where $i(t)$ is the total fraction of the population infectious. In this case, the linearized system is still effectively 2-dimensional with parameters $\gamma$, $\alpha/\gamma$, and $R_{0}$, where $R_{0}$ is given by Eq.(\ref{eq:R0new1}). For this reason we expect our averaging approximation to be {\it exact} in the limit of heterogeneous infectivity only, and a good approximation when the variation in recovery rates is not too large. 

Examples are shown in Fig.\ref{fig:HigherD}, where each panel shows results for an $M\!=\!2$ model in which asymptomatics are significantly more (a) and less (b) infectious than symptomatics\cite{Schwartz2020}. Symptomatic infectives are denoted with the subscript $1$ and asymptomatics with the subscript $2$. The optimal closure period is plotted versus the fraction of asymptomatics, $p_{2}$. Within each panel the different colors correspond to no variation in recovery rates (red), moderate variation (blue), and large variation (green). Simulation determined $T_{\text{min}}$ are shown with points and predictions from the averaging theory shown with solid lines. The initial conditions for simulations follow the SEIR model convention-- parallel to the principal solution of Eq.(\ref{eq:Floquet}), $\boldsymbol{\Psi}^{p}(0)$ -- except that the fraction in each infectious class is $i_{m}(0)\!=\!p_{m}[\boldsymbol{\Psi}^{p}]_{2}$. The model parameters were chosen to match similar models\cite{Covasim,Schwartz2020}, which were fit to multiple COVID-19 data sources. As expected, the agreement between theory and simulations ranges from excellent to fair depending on the heterogeneity in recovery rates
\begin{figure}[t]
\vspace{0.2cm}
\center{\includegraphics[scale=0.235]{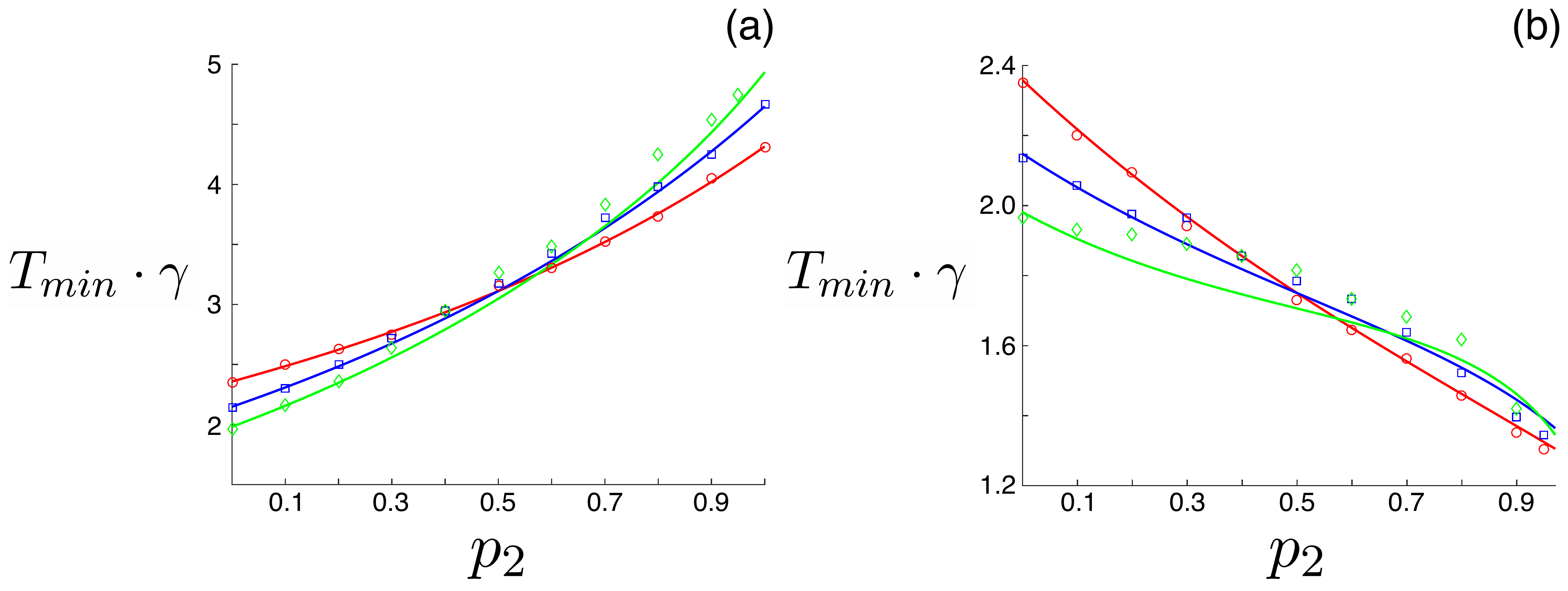}}
\caption{Optimal closure period for a heterogeneous SEIR model with symptomatic and asymptomatic infection as a function of the fraction of asymptomatics. (a) Increased infectivity for asymptomatics, $\beta_{1}\!=\!2.1\cdot\gamma_{1}$ and $\beta_{2}\!=\!2.6\cdot\gamma_{2}$. The solid lines are theoretical predictions and the points are simulation-determined minima for initial fractions of non-susceptibles $10^{-5}$. Each series has different recovery times: red ($\gamma_{1}^{-1}\!=\!10\!\cdot\!\text{days}$, $\gamma_{2}^{-1}\!=\!10\!\cdot\!\text{days}$), blue ($\gamma_{1}^{-1}\!=\!12\!\cdot\!\text{days}$, $\gamma_{2}^{-1}\!=\!8\!\cdot\!\text{days}$), and green ($\gamma_{1}^{-1}\!=\!14\!\cdot\!\text{days}$, $\gamma_{2}^{-1}\!=\!7\!\cdot\!\text{days}$). The incubation period is $\alpha^{-1}\!=\!7\!\cdot\!\text{days}$. (b) Decreased infectivity for asymptomatics. Model parameters are identical to (a) except $\beta_{2}\!=\!1.5\cdot\gamma_{2}$.}
\label{fig:HigherD}
\end{figure}
 
\section{Discussion}
Figure \ref{fig:HigherD} demonstrates that the optimal closure period for COVID-19 can depend significantly on the amount of asymptomatic spread, particularly if there is a large difference in infection rates compared to symptomatic cases. Since asymptomatic spread is difficult to measure directly, especially in the early stages of an emerging disease outbreak, it may be difficult to estimate the optimal control accurately enough for periodic closure to be an actionable strategy on its own. A possible solution is to deploy effective and widespread testing within a population, {\it early}, and capture the fraction of asymptomatic infections. In any case, if basic parameters are known for an emerging disease dynamics, periodic closure is very effective -- producing large reductions in the final outbreak size (e.g., Fig.\ref{fig:TimeSeries}(b))-- and can be predicted using our methods. 

\textcolor{black}{An additional component of population heterogeneity not treated in this work is age dependence, 
which is known to be particularly important for modelling the COVID-19 pandemic. When considering expanded models that include age
compartments, various mixing mechanisms across age groups generate different reproductive rates of infection~\cite{Blyuss_2020,Hilton_2020,Wilder_2020}.
One extreme compartmented grouping is to decompose a population into
young, middle aged, and seniors with age-dependent contact rates between groups, age-dependent recovery periods, and some modest age-dependence in incubation periods. Under weak inter-age mixing assumptions, the result is a system of equations similar to Eqs.(\ref{eq:SEIRnew1}-\ref{eq:SEIRnew2}). As demonstrated in Sec.\ref{sec:Hetero}, the emergence of an optimal periodic control depends primarily on $R_{0}$ and the mean incubation period, and persists in spite of population heterogeneity. Although our controls are based on mean epidemiological parameters, it is easy to see how such controls may be distributed across age-dependent groups, and/or spatial clusters. Thus, we expect the inclusion of age-dependent effects to quantitatively change the results presented, but leave our methodology and qualitative findings intact.}

\textcolor{black}{Finally, we should remark that in addition to the heterogeneity discussed,
parameter fluctuations for COVID-19 spread can occur in space and time.
In fact, noise in reporting, differences in local policies, and adherence to the various forms of
intervention may cause drastic fluctuations in the local spreading parameters. Given these facts, 
the well-mixed nature of our model may be insufficient to provide accurate optimal-control predictions. In such cases, a
meta-population or network framework may be more appropriate. Yet, the approach that we lay out can be naturally generalized to more accurate and heterogeneous contact-network models, particularly since SIR and SEIR model dynamics on random networks can be described by relatively low-dimensional dynamical systems\cite{Volz2,Hindes1,Zhao_SEIR,schwartz1992small}, which could be analyzed using the methods described in Sec.\ref{sec:methods}. For small levels of infection, the main contribution from contact heterogeneity is to increase the effective, network $R_{0}$. Once the correct $R_{0}$ is assumed, however, we expect the network results to be similar to those presented here, though this is a subject for future study.}
 
%Exploring the role of network effects represents an interesting avenue for future research, but is beyond the scope of this paper.  

\section{Conclusion}
In conclusion, a main socio-economic issue with an emerging virus, in the absence of
vaccines and treatments, is the enormous damage at all levels of
a population. Here we considered a simple approach to model and control an emerging virus
outbreak with a finite incubation period. We show that by tuning periodic
control of social contact rates, there exists an optimal period that
naturally minimizes the outbreak size of the disease, as long as the reproductive number is 
below a predictable threshold and there is not a time-scale separation 
between incubation and recovery. Our basic assumption for the
existence of such an optimal control rests on early detection of the disease,
in which non-susceptible populations are small. Such a basic assumption allows one to analytically
predict the optimal period, and provide parameter regions in which an optimal control exists. While in general it has been suggested that periodic closure may help curb the spread of an infectious disease like COVID-19, the implementation of such measures has been, to the best of our knowledge, mostly based on observations of recovery periods and absence of new cases for a given period of time. In this paper, we provide a general formulation that can be utilized to rationally design optimal intervention release protocols. While we start from an SEIR model and expand to heterogeneous models that capture the basic dynamics of COVID-19, our theory can be generally applied to acute infections, with the caveat that recovery and incubation periods should be roughly on the same time scale.

\section*{Supporting information}
% Include only the SI item label in the paragraph heading. Use the \nameref{label} command to cite SI items in the text.
\paragraph*{S1 Appendix.}\label{S1_Appendix}
{\bf Optimal control analysis.} Supporting calculations and derivation of the outbreak-minimizing periodic control for the SEIR model. Additional simulation and analysis for both smooth and asymmetric control. 

\section*{Acknowledgments}
The authors acknowledge useful discussions with the IBM COVID19 modeling taskforce. 
%.
\nolinenumbers

% Either type in your references using
% \begin{thebibliography}{}
% Text
% \end{thebibliography}
%
% or
%
% Compile your BiBTeX database using our plos2015.bst
% style file and paste the contents of your .bbl file
% here. See http://journals.plos.org/plosone/s/latex for 
%\bibliography{Newcites}

\begin{thebibliography}{10}
\bibitem{Floornote}
The floor function rounds its argument down to the nearest integer while
  $\text{mod}(x,2)$ denotes the remainder of integer-division of $x$ by $2$;.
\end{thebibliography}

\end{nolinenumbers}
\end{document}